\documentclass[aps,prb,superscriptaddress,twocolumn,showkeys,amsmath,amssymb,floatfix]{revtex4-2}
\usepackage{amsmath,amssymb,bbm,mathrsfs,bm,braket,color,graphicx,comment,amsfonts,dsfont}
\usepackage[colorlinks,citecolor=blue,urlcolor=blue]{hyperref}

\begin{document}

\title{Quantum oscillations reveal sixfold fermions in cubic $\beta$-PtBi$_2$}

\author{E. F. Bavaro}
\affiliation{Centro At\'omico Bariloche and Instituto Balseiro, U. N. de Cuyo, 8400 San Carlos de Bariloche, Argentina}
\author{J. Castro}
\affiliation{Departamento de F\'{i}sica de la Materia Condensada, GIyA-CNEA, Av. General Paz 1499, (1650) San Mart\'{i}n, Pcia. de Buenos Aires, Argentina}
\affiliation{Instituto de Nanociencia y Nanotecnolog\'{i}a (INN CNEA-CONICET), 1650 San Mart\'{i}n, Argentina}
\author{V. Vildosola}
\affiliation{Departamento de F\'{i}sica de la Materia Condensada, GIyA-CNEA, Av. General Paz 1499, (1650) San Mart\'{i}n, Pcia. de Buenos Aires, Argentina}
\affiliation{Instituto de Nanociencia y Nanotecnolog\'{i}a (INN CNEA-CONICET), 1650 San Mart\'{i}n, Argentina}
\author{J. I. Facio}
\affiliation{Centro At\'omico Bariloche and Instituto Balseiro, CNEA, CONICET and U. N. de Cuyo, 8400 San Carlos de Bariloche, Argentina}
\affiliation{Instituto de Nanociencia y Nanotecnologia (CNEA-CONICET), Av. Bustillo, 9500, Argentina}
\author{V. F. Correa}
\affiliation{Centro At\'omico Bariloche and Instituto Balseiro, CNEA, CONICET and U. N. de Cuyo, 8400 San Carlos de Bariloche, Argentina}

\date{\today}

\begin{abstract}
We report a study of de Haas–van Alphen oscillations in high-quality single crystals of cubic $\beta$-PtBi$_2$. In combination with density functional theory calculations, we identify quantum oscillations associated with all Fermi surface sheets predicted by theory. Our results uncover three small electron pockets centered on a sixfold band-touching point located approximately 25\,meV below the Fermi level at the $R$ point of the Brillouin zone. These findings firmly establish the presence of sixfold fermions in close proximity to the Fermi energy of $\beta$-PtBi$_2$.
\end{abstract}

\maketitle

\section{Introduction}

Crystalline symmetries play a central role in shaping the nature of electronic excitations in solids. This has been exemplified by studies investigating the allowed types of energy band crossings in different space groups \cite{Bradlyn2016,Wieder2016}. A remarkable outcome of this effort is the realization that condensed matter systems can host generalized analogues of fundamental particles such as Dirac and Weyl fermions. These include, for instance, threefold-, fourfold-, sixfold-, and eightfold-degenerate fermions, which can arise at high-symmetry points of the Brillouin zone in materials with specific space group symmetries.

Accordingly, there is an active search for materials in which such crossings occur in close proximity to the Fermi energy \cite{Tang2017,Chang2017,Yang2017,sanchez2019,Takane2019,schroter2019,rao2019,Rappe2019,Niels2020,Yang2020,Jin2021,Guo2021,Jin2022,Ju2022}. In this work, we focus on cubic $\beta$-PtBi$_2$, a well-studied semimetal known for its outstanding electronic properties. These include a large residual resistivity ratio (RRR $=\rho(300\,\text{K})/\rho(0) > 500$, with $\rho_0 < 0.1\,\mu\Omega$cm), a long mean free path ($l_0 \sim 1\,\mu$m below $T = 10\,$K), and an extremely large, non-saturating transverse magnetoresistance (MR $= [R(H) - R(0)] / R(0) \sim 10^5$ at $T = 1.8\,$K and $\mu_0 H = 33\,$T) \cite{Gao2017,Correa2022}.
 
In addition, $\beta$-PtBi$_2$ crystallizes in the cubic space group SG 205, which allows for the presence of stable triple Dirac points at the $R$ points of the Brillouin zone, protected by threefold rotational symmetry \cite{Bradlyn2016}. Angle-resolved photoemission spectroscopy (ARPES) experiments have recently indicated the observation of such sixfold fermions, the triple Dirac point lying only 20\,meV below the Fermi energy \cite{Thirupathaiah2021}. 
  Based on this finding, it is natural to expect signatures of these states in bulk properties of $\beta$-PtBi$_2$. 
  Quantum oscillations and, particularly, their macroscopic manifestation in magnetization (de Haas-van Alphen effect, dHvA) and electrical resistivity (Shubnikov-de Haas effect, SdH) 
  are a natural choice for probing the bulk Fermi surface of metals and semimetals \cite{Shoenberg1984,Zhu2015,Fauque2018,Hu2019,Zhao2022}. 
  However, previous quantum oscillation experiments in $\beta$-PtBi$_2$  have detected only Fermi surfaces significantly larger than those expected for the sixfold fermions inferred from ARPES.

Specifically, Gao \textit{et al.} \cite{Gao2017,Gao2019} reported a hole pocket ($\alpha$) along the $\Gamma$–$M$ line of the Brillouin zone and an electron pocket ($\beta$) centered at the $R$ point. Both frequencies are well reproduced by density functional theory (DFT) calculations, which indicate that the latter pocket does not originate from a sixfold-degenerate band. Zhao \textit{et al.} \cite{Zhao2018} observed similar oscillation frequencies but also proposed the presence of two larger electron pockets centered at $\Gamma$. These discrepancies may arise from differences in sample quality and/or variations in magnetic field orientation and intensity. In all cases, the observed pockets exhibit small effective masses, consistent with the semimetallic nature of the material.

Here, we perform high-resolution dHvA experiments on high-quality single crystals of $\beta$-PtBi$_2$. Fast Fourier transform (FFT) spectra of the oscillations reveal at least four set of frequencies. Two of them are in good agreement with previously reported data \cite{Gao2017,Zhao2018,Gao2019} and according to our DFT calculations correspond to the $\alpha$ hole pocket ($f_{\alpha} \approx$ 280\,T) and the $\beta$ electron pocket ($f_{\beta} \approx$ 710\,T). We also find a larger frequency ($f_{\delta}\approx$ 1100\,T), which can be assigned to an electron pocket centered at $\Gamma$. 
Remarkably, we observe an additional set of peaks at low frequencies ($f_{\nu} < 100$\,T). A detailed comparison with DFT reveals that their angular dependence is consistent with small electron pockets originating from a sixfold band-touching point at the $R$ point. These findings are fully consistent with the ARPES data of Ref.~\cite{Thirupathaiah2021} and provide compelling bulk evidence for the presence of multifold fermions at the Fermi level of $\beta$-PtBi$_2$.

The rest of this paper is organized as follows. 
Section \ref{sec_methods} describes experimental and theoretical methods used in this work. 
Section \ref{sec_results} describes the experimental results together with DFT calculations. 
Section \ref{sec_discussion} contains a discussion of the interpretation of our observations while 
Section \ref{sec_conclusions} presents our concluding remarks. 
A detailed analysis of the smallest frequencies using the Lifshitz–Kosevich (LK) theory is given in Appendix \ref{app:dHvA}.
Appendix \ref{app:spurious} contains experimental results aimed at discarding spurious non-intrinsic contribution to the dHvA oscillations.  
Lastly, the Appendix \ref{app:dft} provides details on the dependence of the DFT results on various technical choices.  

\section{Methods}
\label{sec_methods}

\subsection{Experiments}
\begin{figure}[t]
 \centering
\includegraphics[width=0.9\linewidth]{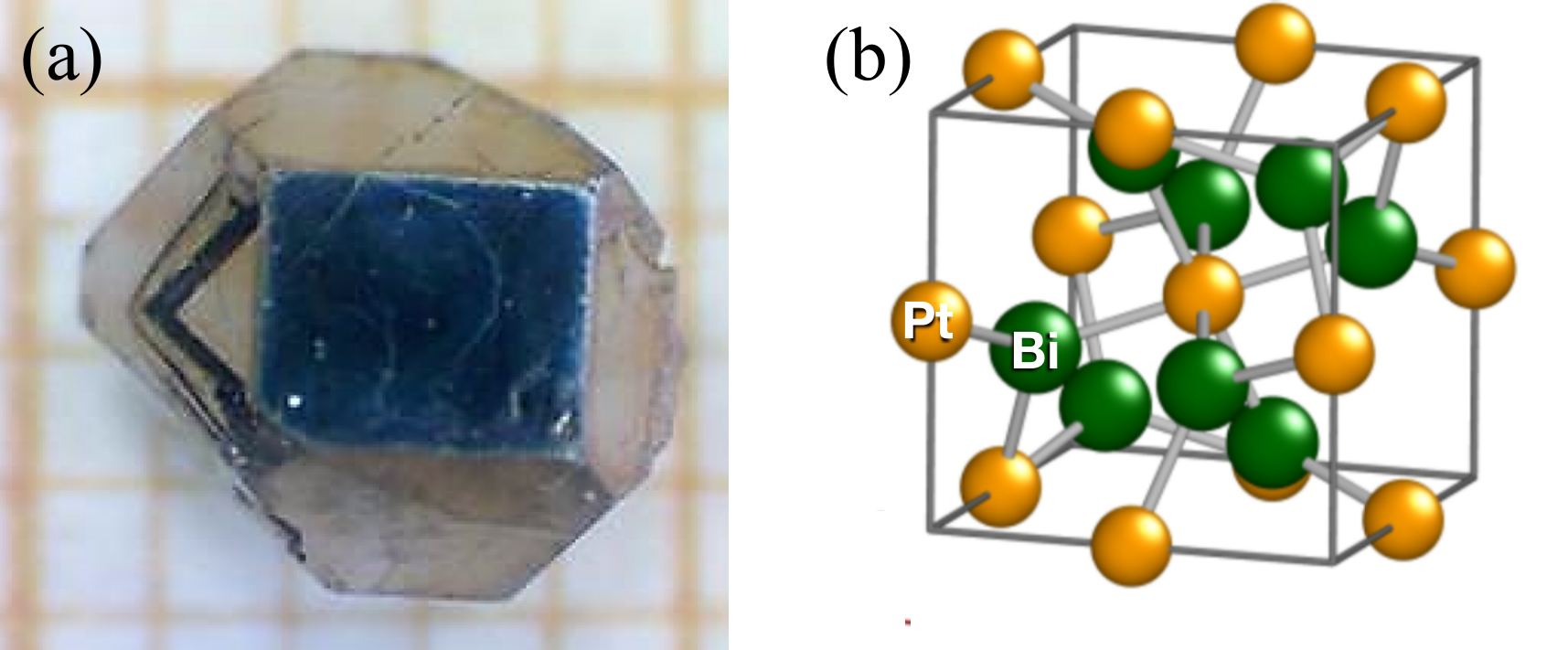}
\includegraphics[width=\linewidth]{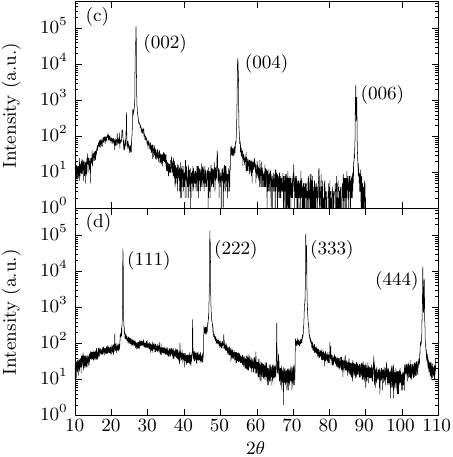} 
 \caption{(a) Sample of $\beta$-PtBi$_2$. (b) Crystal structure in the conventional cubic cell representation. (c,d) X-ray diffraction patterns corresponding to the $\left( 0\,0\,l \right)$ and $\left( k\,k\,k\, \right)$ families of planes.}
\label{fig:xrd} 
\end{figure}

High-quality single crystals (Fig. \ref{fig:xrd}a) were grown by the self-flux technique as described elsewhere \cite{Correa2022}. Two large crystals were separated for the dHvA experiments. 
X-ray diffraction (XRD) experiments confirm the correct cubic structure.
Fig. \ref{fig:xrd}c) shows Cu-$K_{\alpha}$ $\theta - 2\theta$   XRD scans corresponding to the family of planes $\left( 0\,0\,l \right)$ and Fig. \ref{fig:xrd}d) for $\left( k\,k\,k\, \right)$. No spurious peaks other than Cu-$K_{\beta}$ reflections are observed. 

dHvA experiments for selected directions of the applied magnetic field were performed in two squid magnetometers: a 5-Tesla Quantum Design MPMS5 and a 7-Tesla Cryogenic S700X. 
In Ref.~\cite{bavaro_2025_15857139}, representative raw data is provided.

\subsection{Theory}
The main theoretical results of this work are based on DFT calculations performed using the FPLO code (v.22) \cite{Klaus1999}. 
We utilized the generalized gradient approximation (GGA) \cite{Perdew1996} and incorporated spin-orbit coupling within the fully-relativistic four-component formalism implemented in FPLO. Brillouin zone integrations were carried out using a tetrahedron method with a mesh consisting of $12\times12\times12$ subdivisions. 
 We use an enhanced basis function set, built as in Ref. \cite{lejaeghere2016}. Overall, the enhancement of the basis has a minor effect on the resulting band structure, but it affects the position of the triple Dirac point relevant in this work (in the order of 20\,meV in comparison with the default FPLO basis).

We use the cubic experimental crystal structure (SG 205) with lattice parameter $a=6.6954\,$\AA~\cite{Correa2022}.
The structure is shown in Figure \ref{fig:xrd}b  and has a Pt atom at the position (0.5,0,0) (in units of the lattice vectors) and a Bi atom at (0.12913,0.12913,0.12913).
In the Appendix \ref{app:dft}, we describe how our results are affected when considering other methodologies, such as a different exchange and correlation functional or a fully relaxed crystal structure.

For the calculation of the quantum oscillation spectra, we use the dHvA module of FPLO, which searches for extremal orbits starting from a calculated Fermi surface. The latter was obtained via an adaptive-mesh algorithm initiated with a $k$-mesh having $18\times18\times18$ subdivisions and adaptively improved over four refinement cycles. For this calculation, we use a Wannier Hamiltonian based on Bi $6s$ and $6p$ orbitals, as well as Pt $6s$ and $5d$ orbitals and built with the symmetry-conserving projection method developed in Ref. \cite{Koepernik2023}. The resulting band structure deviates from the DFT results in the order of 1\,meV.

\section{Results}
\label{sec_results}

 \begin{figure}[t!]
 \centering
\includegraphics[width=\linewidth]{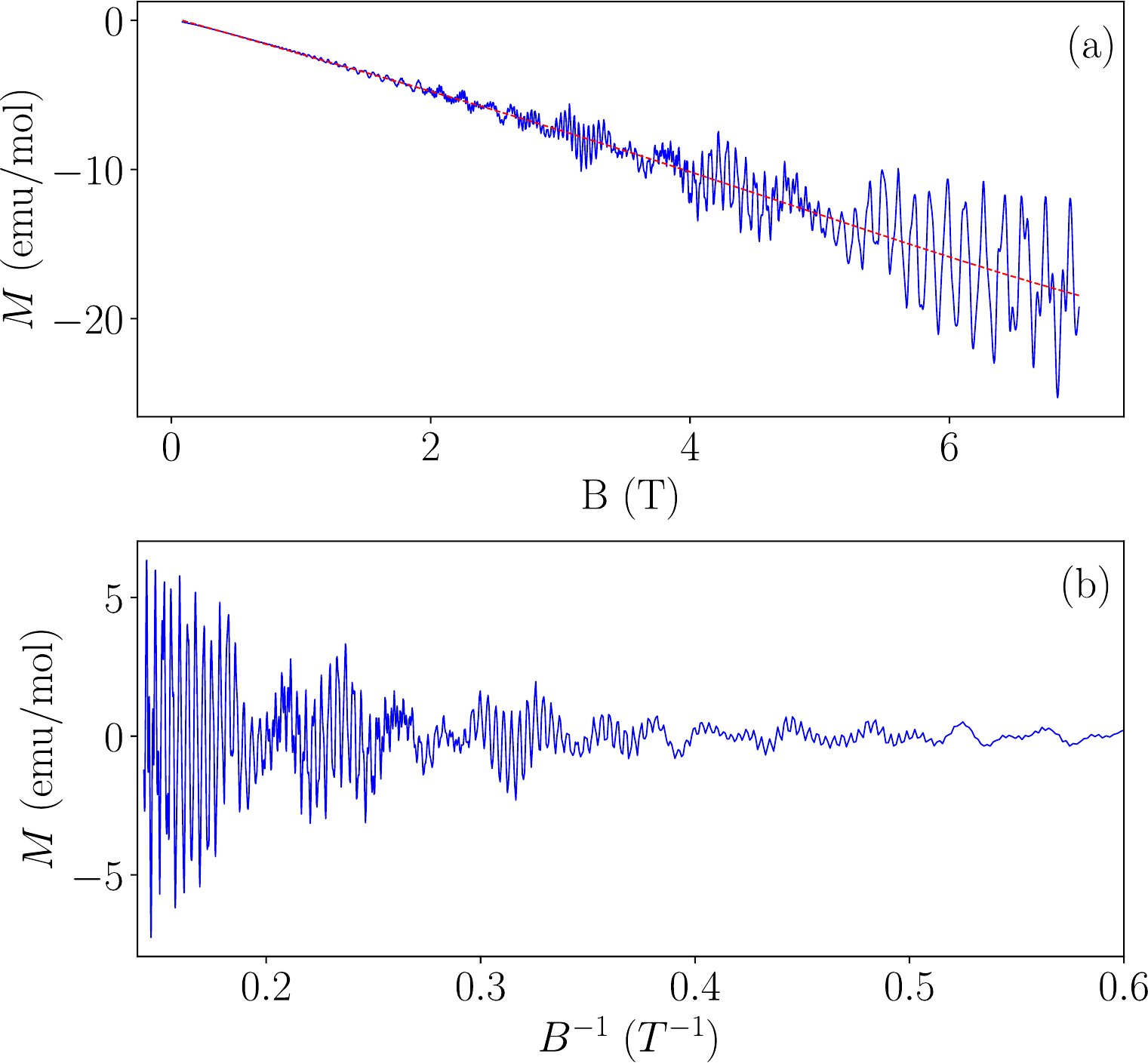} 
 \caption{Magnetization as a function of applied magnetic field (a) and of the inverse of the magnetic field (b). The dashed line in (a) corresponds to the basal line subtracted for the study of the quantum oscillations.
 }
\label{fig:exp} 
\end{figure}

Fig. \ref{fig:exp}a shows the longitudinal magnetization as a function of a magnetic field applied along the $[111]$ direction at $T=2.1\,$K. The response is diamagnetic, and clear oscillations can be observed above $B=1\,$T. 
In order to analyze the frequency spectrum of these oscillations, we subtract the smooth diamagnetic background. Fig. \ref{fig:exp}b exhibits the resulting oscillating component of the magnetization as a function of $1/B$.

 \begin{figure*}[t]
 \centering
\includegraphics[width=\linewidth]{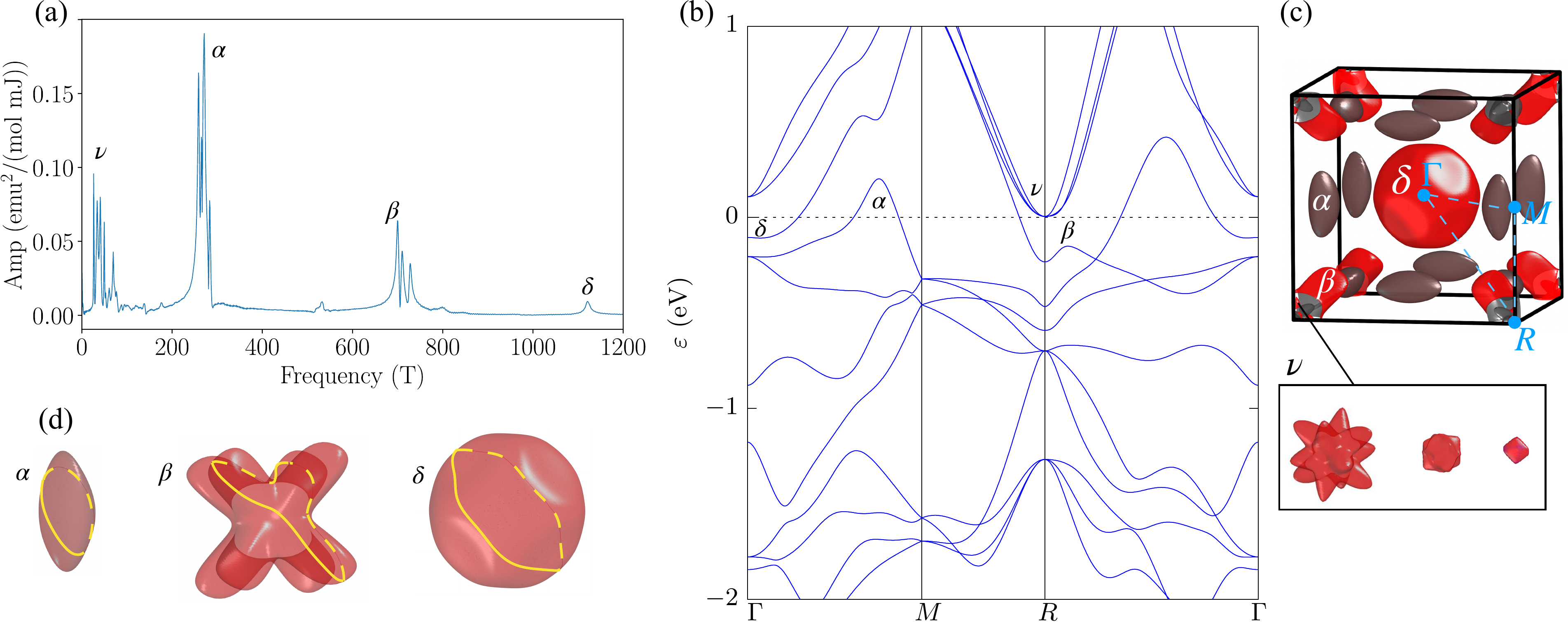} 
 \caption{(a) Frequency spectrum of the oscillations for the magnetic field applied along the [111] direction. (b) DFT band structure based on PBE+SOC. (c) DFT Fermi surfaces. The pockets $\alpha$, $\beta$ and $\delta$ correspond to the Fermi energy as obtained from DFT. The bottom panel shows a zoom of the small pockets $\nu$, which are obtained by placing the Fermi energy 25\,meV above the triple Dirac point. (d) Representative theoretical extremal orbits (yellow lines) of the $\alpha$, $\beta$ and $\delta$ pockets for the field along the [111] direction. Notice that there is another extremal orbit arising from symmetry-related $\alpha$ pockets and two additional extremal orbits of the $\beta$ pocket.
 }
\label{fig:dft} 
\end{figure*}

Fig. \ref{fig:dft}a presents the corresponding frequency spectrum. A second sample from a different batch shows the same spectrum. 
Among the peaks of larger amplitude, those around 260\,T and 700\,T are in good agreement with previous experimental works. The former has been associated with the Fermi surface pockets $\alpha$, which lie along the $\Gamma$-$M$ lines and it is also accompanied by a second harmonic at $\sim$520\,T.
 The structures around 700\,T originate from the $\beta$ electron pocket centered at $R$.
 There is also a small peak around 1100\,T which may correspond to the $\delta$ electron pocket observed in previous SdH experiments \cite{Zhao2018}. 
 The main experimental observation of this work, previously not reported to the best of our knowledge, is the series of peaks below 100\,T, which arise from pockets formed by a sixfold band touching at the $R$ point as we argue below.

 Fig. \ref{fig:dft}b presents the band structure as obtained from full-potential, fully relativistic DFT calculations. The associated Fermi surface (Fig. \ref{fig:dft}c) is composed of the $\alpha$ and $\beta$  pockets discussed above, and of an electron pocket ($\delta$) centered at $\Gamma$. The latter yields the peak around 1100\,T. 
 The $\alpha$, $\beta$ and $\delta$ pockets found by our DFT calculations match all observed frequencies larger than 100\,T (Table \ref{table})~\footnote{In our measurements, four distinct frequencies are attributed to the $\alpha$ pocket, whereas DFT calculations predict only two. This discrepancy can be accounted for by a slight misalignment of the magnetic field with respect to the (111) direction (on the order of 1$^{\circ}$), which—according to our calculations for various field orientations, such as (112) or (110)—leads to a splitting of the extremal orbits of the $\alpha$ pocket into pairs. In contrast, according to the calculations and consistent with the observations, such splitting does not occur for the orbits of the $\beta$ and $\delta$ pockets. }. The largest disagreement is obtained for the $\beta$ pocket, the predicted frequencies being between 11\% and 17\% larger than observed.

 In addition, a sixfold band touching point at $R$ lies almost at the Fermi energy. 
 As discussed in the Appendix \ref{app:dft}, its exact energy is sensitive to choices such as the basis set used with FPLO or the exchange and correlation functional. 
 For example,  Ref. \cite{Gibson2015} finds that calculations based on the modified Becke–Johnson functional yield the band crossing point below the Fermi energy.
We find that placing the Fermi energy 25\,meV above the sixfold band touching point yields three additional small electron pockets (enlarged in Fig. \ref{fig:dft}c).
 The corresponding extremal-orbit frequencies match fairly well the observed spectrum below 100\,T (Table \ref{table}). Notably, a rigid shift of the Fermi energy of this amount significantly worsens the level of agreement between theory and experiment of the remaining frequencies, suggesting that the discrepancy should not be attributed to doping, but rather to small relative shifts between different bands in the calculated electronic structure and the actual material.

\begin{table}[t]
\begin{tabular}{|c|c|c|}
\hline
Fermi surface & Frequency (T) (Exp.) & Frequency (T) (DFT) \\ \hline
$\nu$      & 26.3, 34.2, 41.3, 49.5, 69.8                   & 25, 48, 75 (*)                 \\ \hline
$\alpha$      & 259, 264, 270, 283.5                  & 284, 297                 \\ \hline
$\beta$       & 700, 710, 728                  & 781, 792, 844                 \\ \hline
$\delta$      & 1121                 & 1075                \\ \hline
\end{tabular}
\caption{Observed and calculated quantum oscillation frequencies. 
The frequencies indicated with (*) correspond to a calculation where the Fermi energy has been shifted by 25\,meV above the triple Dirac point. }
\label{table}
\end{table}

 \begin{figure}[t]
 \centering
\includegraphics[width=\linewidth]{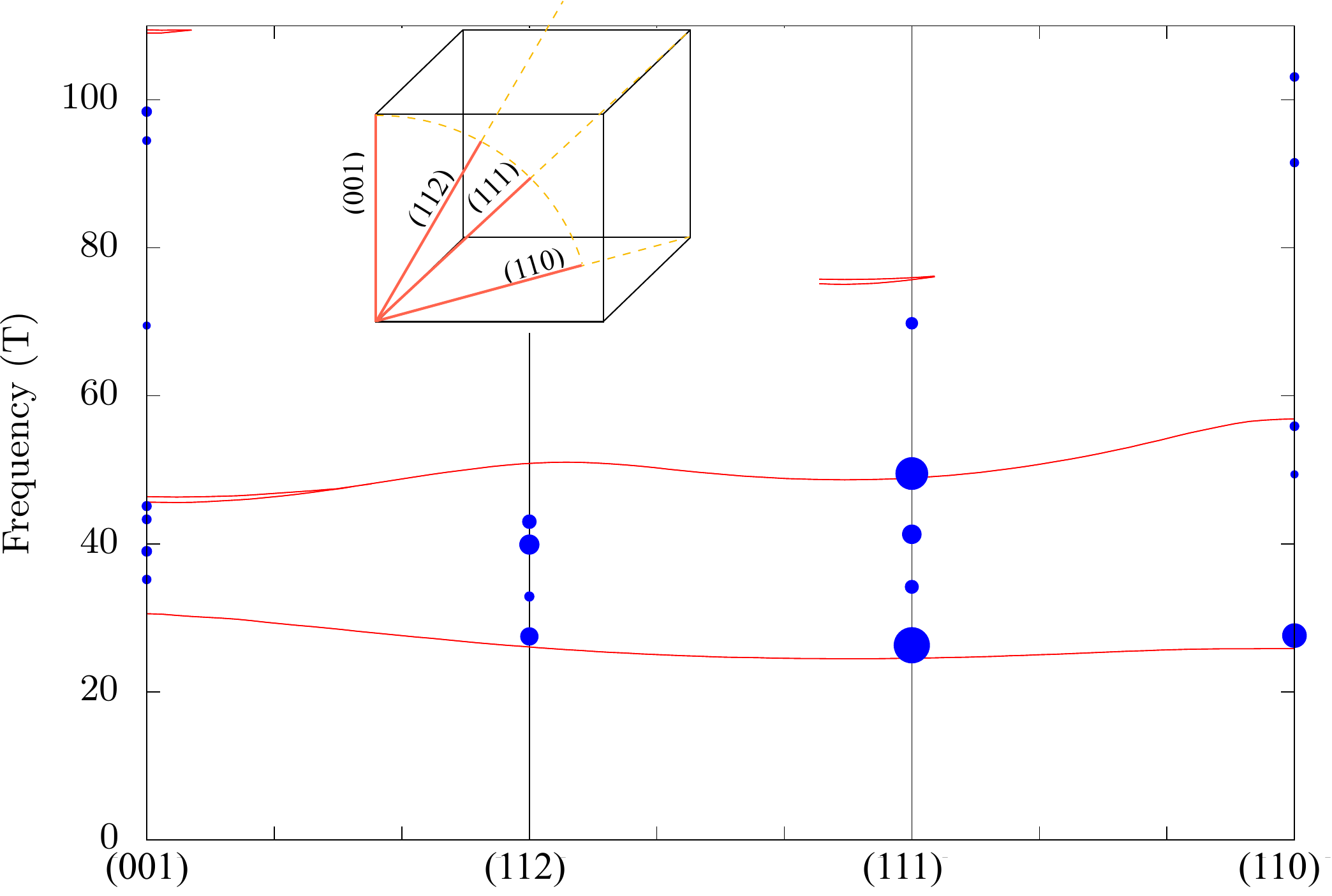} 
 \caption{Angular dependence of the low-frequency quantum oscillations. Blue dots correspond to the experimental results and continuous lines to the DFT calculation. The size of the dots increases according to the observed amplitude. The theoretical results originate from the $\nu$ pocket centered at $R$, which was constructed by placing the Fermi energy 25\,meV above the band touching point. \textit{Insert:} Sketch of the directions at which the magnetic field has been applied.
 }
\label{fig:ang} 
\end{figure}
 
 Fig. \ref{fig:ang} shows the angular dependence of these low-lying frequencies in comparison with the DFT calculation. The experimental data correspond to four discrete directions, all of them lying on the same crystallographic plane. The overall agreement supports the interpretation of these low-frequency quantum oscillations as evidence that the sixfold band touching point lies approximately 20\,meV below the Fermi energy.
 This conclusion is in agreement with a similar finding based on ARPES experiments  \cite{Thirupathaiah2021}.

Fitting the temperature dependence of the oscillation amplitude using the Lifshitz-Kosevich (LK) formula, we obtain effective cyclotron masses associated with the five smallest frequencies observed for the magnetic field along the $(111)$ direction, as described in the Appendix \ref{app:dHvA}. In order of increasing frequency, we obtain $m_\nu/m_e = 0.07 \pm 0.02, 0.1 \pm 0.02, 0.11 \pm 0.02, 0.11 \pm 0.02$ and $0.20\pm 0.04$.  These rather small values are in line with the calculated cyclotron masses, which lie between 0.11 and 0.26.

Lastly, it is apparent that the FFT results in Fig. \ref{fig:ang} contain more peaks than expected from the DFT calculation. 
Some of the additional peaks can be explained within errorbars as higher harmonics of fundamental oscillation modes, but others cannot be attributed to such effects.
It can be the case that the DFT calculation does not capture fine details of the actual Fermi surface, leading to the omission of certain extremal orbits. One possible source of such fine details may be the splitting between Kramers pairs due to the applied magnetic field. Namely, in the calculation the presence of time-reversal symmetry (together with inversion symmetry) forces all the bands to be doubly degenerate. The breaking of time-reversal symmetry by the applied field could, in principle, enrich the frequency spectrum. Other effects of the magnetic field  associated with proximity to the quantum limit may also be at play.
Further studies at higher magnetic fields would help to clarify this issue.

\section{Discussion}
\label{sec_discussion}
It is natural to question the origin of the discrepancies between different studies based on quantum oscillations in $\beta$-PtBi$_2$. 
Regarding differences between SdH and dHvA experiments, extremal areas and effective masses extracted from the former \cite{Gao2017} are a little larger than those obtained from the latter \cite{Gao2019}, even for samples from the same group. 
Varying results between the two effects are not surprising, even though both are ultimately related to oscillations in the density of states. The key difference lies in the fact that the dHvA effect reflects an equilibrium property derived from the free energy, whereas the SdH effect is “\emph{contaminated}” by non-equilibrium mechanisms such as the scattering rate of electrons near the Fermi surface.
In this sense, the dHvA effect usually provides more accurate results than SdH measurements \cite{Shoenberg1984}. This effect may be especially pronounced in semimetals, particularly $\beta$-PtBi$_2$, due to the extreme magnetoresistance which greatly masks the resistance oscillations \cite{Gao2017,Zhao2018}. By contrast, magnetization oscillations are comparable to the steady diamagnetic background even at moderate fields \cite{Gao2019}.

Regarding differences between different dHvA experiments, on general grounds it is reasonable to suspect that differences arise from the quality and characteristics of the samples. Indeed, due to the close proximity of the sixfold band touching to the Fermi energy, it is plausible that accidental doping plays a role. Our calculations, nevertheless, suggest that a scenario in which the observation of the $\nu$ pockets in our experiments stems from our samples being electron doped is unlikely because the required shift of the Fermi energy significantly affects the remaining oscillation frequencies.
Namely, for the Fermi surface obtained with the Fermi energy 25\,meV above the triple Dirac point, the agreement between DFT and the observed frequencies of the $\alpha$, $\beta$ and $\delta$ pockets worsens by $\sim$20\%.

Given the small frequencies and effective masses associated with the sixfold fermions, one might also consider possible contributions from spurious phases characterized by small Fermi surface pockets, typically residual Bi from the growth process, or the polymorph trigonal $\gamma$-PtBi$_2$.
However, a close examination using different techniques such as energy-dispersive X-ray spectroscopy (EDS), differential scanning calorimetry (DSC) and powder x-ray diffraction (XRD) indicates that this scenario can be ruled out, as shown in Appendix \ref{app:spurious}.

Ultimately, the very reasonable agreement between our observations and DFT calculations as well as with ARPES results \cite{Thirupathaiah2021} lead us to conclude that the interpretation of our experimental results as reflecting the existence of sixfold fermions at the Fermi level is the most natural one.

\section{Conclusions}
\label{sec_conclusions}

We have studied the oscillations of the magnetization as a function of an applied magnetic field in the cubic phase of PtBi$_2$. Our results can be quantitatively interpreted based on fully-relativistic density-functional calculations. The key finding is the unambiguous identification of quantum oscillations originating from sixfold fermions, protected by the threefold rotational symmetry and located in close proximity to the Fermi level. These results bridge a gap in the existing literature, which reported signatures of these multifold fermions via ARPES measurements but lacked corresponding evidence from quantum oscillation experiments. In this way, our study firmly establishes PtBi$_2$ as a material hosting low-energy multifold quasiparticles.

\section{Acknowledgments}

Work partially supported by Agencia Nacional de
Promoción Científica y Tecnológica (ANPCyT) grants number PICT 2019/00371, PICT 2019/02396, PIP 2023-2025 11220220100330CO of CONICET and Universidad Nacional de Cuyo grant number 06/C55-T1.
JIF thanks the Alexander von Humboldt Foundation for support via the Georg Forster Return Fellowship. 

\appendix
\section{Lifshitz–Kosevich analysis}
\label{app:dHvA}

\begin{figure}[t]
 \centering
\includegraphics[width=8cm]{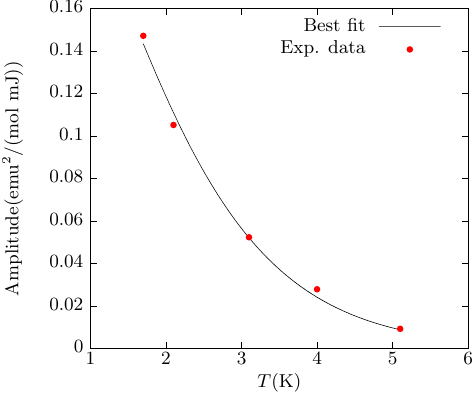} 
 \caption{Temperature dependence of the amplitude of the oscillation of 49.5\,T, for the magnetic field along the direction (111).
 }
\label{app:fits} 
\end{figure}

We fit the oscillation amplitude as a function of frequency using the Lifshitz-Kosevich formula, which reads \cite{lifshitz1956}
\begin{equation}
\Delta M \propto \sqrt{B} R_T R_D \sin(\frac{2\pi f }{B}+\phi).
\label{eqn:L-K}
\end{equation}
Here, $\Delta M$ is the oscillatory part of the magnetization, $f$ the oscillation frequency,   $R_T=\frac{\delta T}{B \sinh(\delta T/B)}$, $R_D = \exp(-\delta T_D / B)$, $\delta=\frac{2\pi^2 k_B m^*}{\hbar e}$ and $\phi$ a phase. $T_D$ is the Dingle temperature which can be written in terms of the relaxation time associated with impurity scattering  as $T_D = \frac{\hbar}{2\pi k_B \tau}$. Effective masses listed in Section \ref{sec_results} are extracted from fits to Equation \ref{eqn:L-K}. A representative fit of the amplitudes using the LK formula is shown in Fig. \ref{app:fits} for $f=49.5$\,T.

The Dingle temperature is extracted from the Dingle plot, namely from the slope of ln$(\Delta M / \sqrt{B} R_T)$ versus $B^{-1}$. 
We obtain $T_D =$ 0.3(1)\,K for $f=26.3$\,T and $T_D =$ 0.6(2)\,K for $f=49.5$\,T.


\section{Non-intrinsic dHvA oscillations}
\label{app:spurious}

\begin{figure}[h]
 \centering
\includegraphics[width=\linewidth]{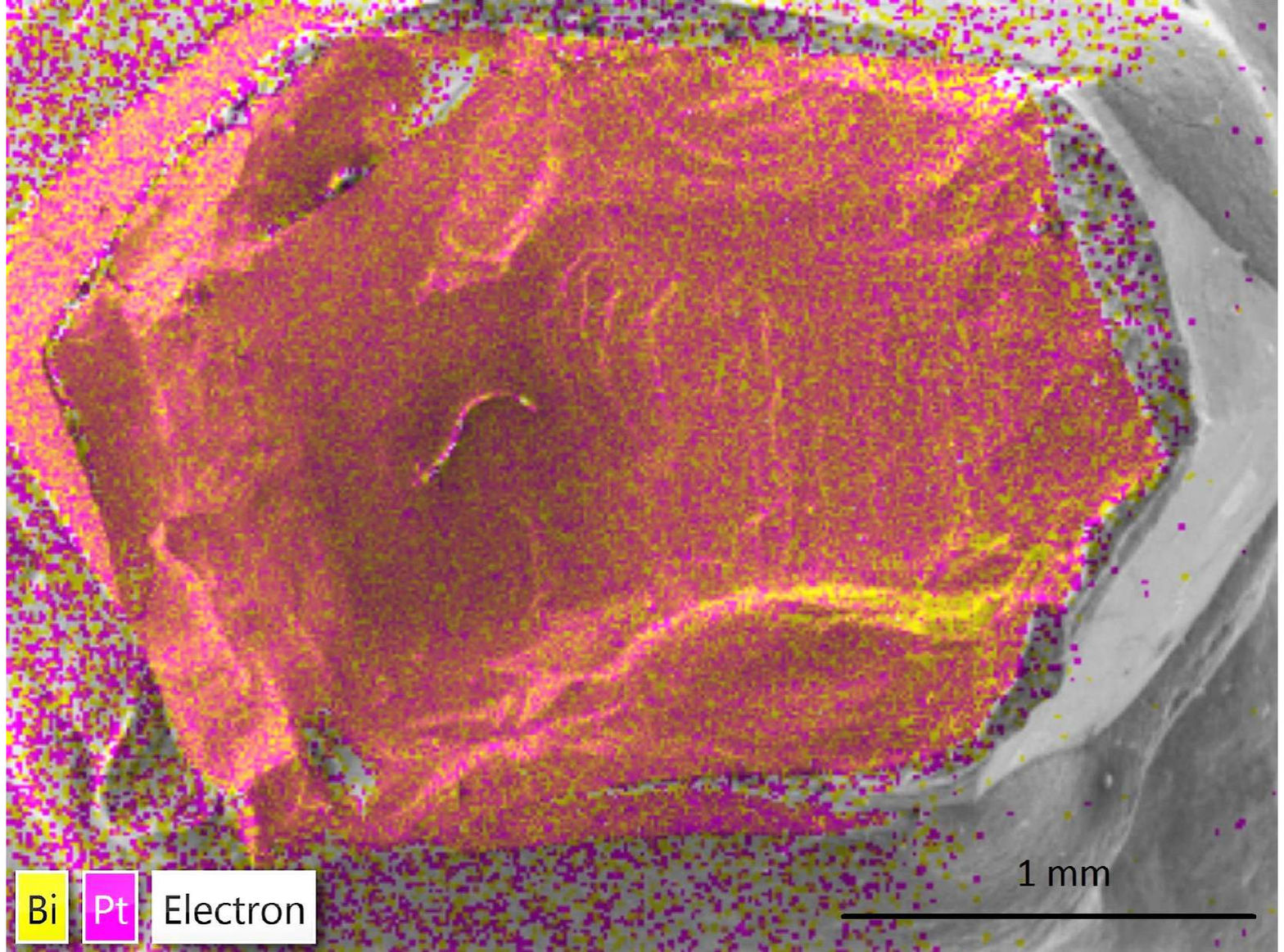} 
 \caption{Energy-dispersive spectroscopy scan mapping the surface of a $\beta$-PtBi$_2$ crystal. Homogeneous Pt and Bi distribution is observed.}
\label{fig:eds} 
\end{figure}

Low-frequency dHvA oscillations ($<$ 100\,T) are known to occur in Bi \cite{Lerner1962} and in the trigonal polymorph $\gamma$-PtBi$_2$ \cite{gao2018,veyrat2023}.
  In order to consider a possible impact by Bi we performed EDS measurements. 
Figure \ref{fig:eds} shows an EDS scan of a crystal taken from the same batch as our first sample. 
The crystal was broken at roughly half pieces aimed at searching for trapped inner Bi flux.
Mapping of Bi and Pt across the newly exposed surface reveals a highly homogeneous distribution, with only a small localized region of elevated Bi content near the bottom-right corner.
Complementary DSC scans puts an uppermost limit to the Bi content of 4\,\%.
Such a low content cannot explain the oscillations amplitude associated with the $\nu$ pockets (Fig. \ref{fig:dft}a).

The same crystal was later cut in very small pieces ($<50\, \mu$m) to perform powder XRD. A very thin scalpel was used. Having in mind the metastable character of $\beta$-PtBi$_2$, we avoided a traditional grinding process based on a mortar in order to prevent local heating that could induce a phase transformation.

\begin{figure}[h]
 \centering
\includegraphics[width=\linewidth]{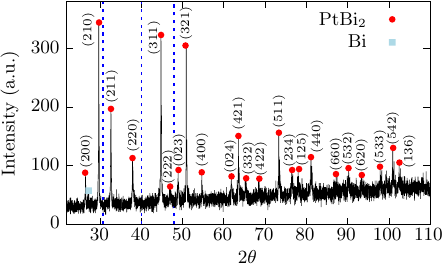} 
 \caption{$\beta$-PtBi$_2$-powder X-ray diffraction spectrum. All the peaks but one (corresponding to Bi, power diffraction file (PDF) 01-085-1329 \cite{gates2019}) are indexed with the correct cubic structure (PDF 01-071-5991). Dashed lines mark the position of the most intense $\gamma$-PtBi$_2$ peaks (PDF 01-084-6757).}
\label{fig:powder_xrd} 
\end{figure}

Figure \ref{fig:powder_xrd} shows the powder XRD
spectrum. Each and every peak, but one, can be indexed with the $\beta$-PtBi$_2$ cubic structure. There is only one small extra peak coming from Bi.
Particularly, no peaks associated with trigonal $\gamma$-PtBi$_2$ are observed.

\section{Dependence of theoretical results on DFT technicalities}
\label{app:dft}

Here we present the dependence of the band touching of six bands at $R$ on different DFT technical choices.
First, we would like to comment on the basis used within FPLO.  The results in the main text were obtained using an enhanced basis built as in \cite{lejaeghere2016}. If the default basis is used, the results are overall similar but the sixfold band touching point moves in energy. In the enhanced basis scheme, this feature places $\sim$4\,meV above the Fermi energy, whereas in the default case, it appears at $\sim$34\,meV. 

We also perform calculations using the projector augmented wave (PAW) method \cite{PAW-1994}, as implemented in the Vienna ab initio package (VASP 6.3.2) \cite{vasp1,vasp2,vasp3} for different exchange and correlation functionals (local-density approximation, LDA\,\cite{LDA-1996}, and GGA) and considering both the experimental and relaxed crystal structures.  We use PAW pseudopotentials that explicitly treat 15 electrons for bismuth ($5d^{10}$, $6s^2$, $6p^3$) and, in the case of platinum we compare the electronic structure for a pseudopotential with 10 electrons (Pt: $5d^9$, $6s^1$) and  16 electrons (Pt-pv: $5p^6$, $5d^9$, $6s^1$). We include the spin-orbit coupling according to the PAW methodology, as described in Ref. \cite{SOC-VASP}.
A 10$\times$10$\times$10 Monkhorst-Pack $k$-point grid for the Brillouin sampling and an plane-wave cut off of 380\,eV are enough for obtaining the desired precision of 1\,meV in the total energy.  The structural relaxations are performed until the forces on each ion are less than 0.01\,eV/Å.

 \begin{figure}[b]
 \centering
\includegraphics[width=0.9\linewidth]{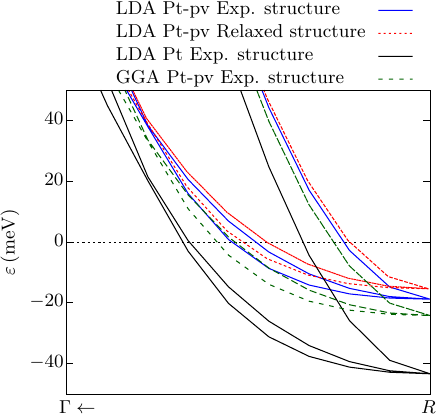} 
 \caption{Band structure near $R$ as computed with VASP with different calculation schemes.}
\label{fig:appbands} 
\end{figure}

We find that the sixfold band at $R$ lies around 20\,meV below the Fermi energy, the exact value depending rather weakly on the use of the experimental crystal structure, or a fully relaxed one, as well as on the use of the LDA or the GGA approximations.
Fig. \ref{fig:appbands} compares the band structure obtained with these different options. We also find a rather strong effect of the precise pseudopotential used: The crossing point moves closer to the Fermi energy as more electrons are considered in the pseudopotential. Specifically, for the experimental crystal structure and the LDA functional, the sixfold band at the $R$ point lies at 43\,meV and 19\,meV below the Fermi energy for the Pt and Pt-pv pseudopotentials, respectively (see Fig. \ref{fig:appbands}).

%


\end{document}